\documentclass{emulateapj}

\usepackage{graphicx}

\usepackage{amssymb}
\usepackage{amsmath}
\usepackage{natbib}
\usepackage[colorlinks=true,linkcolor=blue,anchorcolor=blue,citecolor=blue,urlcolor=blue]{hyperref}  

\shorttitle{Strong gravitational lensing as a tool to investigate the structure of jets at high energies}
\shortauthors{Barnacka et al.}

\begin{document}

\title{Strong gravitational lensing as a tool to investigate the structure of jets at high energies}

\author{Anna Barnacka$^{1,2}$}
\author{Margaret J. Geller$^1$}
\author{Ian P. Dell'antonio$^3$}
\author{Wystan Benbow$^1$}
\affil{$^1$Harvard-Smithsonian Center for Astrophysics, 60 Garden St, MS-20, Cambridge, MA 02138, USA\\
$^2$Astronomical Observatory, Jagiellonian University, Cracow, Poland\\
$^3$Department of Physics, Brown University, Box 1843, Providence, RI 02912}

\email{abarnacka@cfa.harvard.edu}

\begin{abstract}

The components of blazar jets that emit radiation  span a factor of $10^{10}$ in scale.
The spatial structure of these emitting regions depends on  the observed energy.
Photons emitted at different sites cross the lens plane 
at different distances from the mass-weighted center of the lens. 
Thus there are  differences in  magnification ratios and time delays between 
the images of lensed blazars observed at different energies. 
When the lens structure and redshift are known from  optical 
observations, these constraints can elucidate the structure of the source at high energies.
At these energies, current technology is inadequate to resolve these sources
and the observed light curve is thus the sum of the images.
Durations of $\gamma$-ray flares are short compared with typical  time delays;
thus both the magnification ratio and the time delay can be measured for the delayed counterparts. 
These measurements are a basis
for localizing the emitting region along the jet.
To demonstrate the power of strong gravitational lensing,
we build a toy model based on the best studied and the nearest relativistic jet M87.

\end{abstract}


\section{Introduction}

Strong gravitational lensing is a powerful tool for exploring the universe \citep{1992grle.book.....S}. 
It magnifies distant objects and provides a way to observe their structure and detailed properties 
(e.g. \citet{2012ApJ...759...66Y,2012ApJS..199...25P,2012A&A...542L..31L}).  

Blazars are the most luminous energy source in the universe; they are also among the most mysterious.
Strongly lensed blazars offer a new window for elucidating their structure.

The components of blazars that emit radiation from the radio to the $\gamma$-ray span 
a factor of $10^{10}$ in scale. The images of lensed blazars are resolved in the radio and in the optical 
\citep{2013A&A...558A.123M,1992AJ....104.1320O,1995MNRAS.274L...5P,1997MNRAS.289..450K}. 
Their sizes are from sub parsec up to megaparsec.  
Radio interferometry resolves the details of blazar radio emission from the core, the jets, and the extensive lobes 
\citep{2008Natur.452..966M}.
Improved angular resolution of current X-ray satellites demonstrates 
that the X-ray emission from the jet forms structures as large as 
hundreds of kpcs \citep{2006ARA&A..44..463H,2007ApJ...662..900T,2002ApJ...570..543S}.  
At high energies, the technology is inadequate to resolve the sources. 
However, the short variability timescales suggest that the sources of the high energy radiation 
during a flare is of the order of $10^{-3}$~parsec \citep{2011AdSpR..48..998S}. 
It remains unclear whether the radiation source is the same at all energies. 
The source of radiation may be close to the base of the jet or it may originate from blobs moving along the jet at relativistic speed.

Strong lensing provides a potential tool for distinguishing among the possible emission sites. These strongly lensed sources may thus provide fundamental discriminants among
models for high energy emission in blazars.
Photons emitted at the different sites cross the lens plane 
at different distances from the mass-weighted center of the lens. 
Thus the magnification ratios,
and the time delays between the images depend on the location where the radiation originates.  
Because the site of the emission may be energy dependent, the magnifications ratios 
and time delays may differ from one wavelength range to another.
They may even differ from one flare to another in the same source if the radiation originates from knots in the jet.

Here, we develop a method for  using strong gravitational lensing as a tool to constrain
the internal structure of blazar jets. 
We use the detailed observations of M87 to construct a toy model of the source (Section~\ref{sec:M87}). 
We discuss the range of projected physical scales relative to the Einstein radius size (Section~\ref{sec:Scales}).
The first step in the process is the determination of the lens properties 
from the  spatially resolved radio and optical images (Section~\ref{sec:Application}). 
Within the context of the well-constrained $\Lambda$CDM cosmology \citep{2013arXiv1303.5076P},  
distances to the lens and source are also known. 
In $\gamma$-rays, the observed light curve is the sum of the lensed images.
We show that the properties of distinct flares enable retrieval of the properties of the source even though the
images are unresolved. The crucial $\gamma$-ray observations are the amplitude and time delay of the flares (Sections~\ref{sec:VarR} and \ref{sec:VarDT}).

 \begin{figure}
 \includegraphics[width=8.5cm,angle=0]{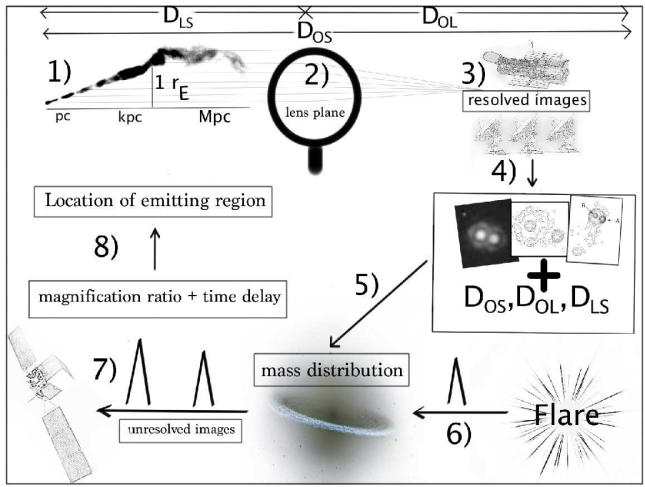}%
 \caption{\label{fig:sketch} Steps in the application of strong gravitational lensing 
                        to unresolved jet structures. 
 		1) Blazar jets extend from sub-parsec up to megaparsec scales. 
		2) A galaxy located close to the  line-of-sight between the source and the observer 
		     acts as a lens with an Einstein radius  of a few kpcs.
		     Radiation emitted in  different regions of the jet crosses the lens plane at different 
		     distances from the center of mass of the lens.
		     Differences in path length result in a different  magnification ratio 
		     and time delay for each  image of lensed blazar.
		3) Observations at radio and optical wavelengths provide resolved images. They also 
		     provide redshifts of the lens and  source. 
		4) The images shows the lensed system B2~0218+35 observed in optical and radio 
		     (Images from: \citep{2000MNRAS.311..389J,1995MNRAS.274L...5P,1999MNRAS.304..349B}).
		5) The resolved images  and the known distances allow  retrieval of the mass distribution of the lens.
		6) The location of emitting region at high energies cannot be resolved with current instruments, 
		    but variability time scales as short as a few hours suggest that the region have to be compact.
		7) The $\gamma$-ray variability timescale is short compare to the time delay. Thus
		     the time delay 
		     and corresponding magnification ratio for the  delayed counterparts of the  radiation source can be estimated.
		8)  Measurement of the time delay and magnification ratio for  the lensed flare 
		     can be used to limit the location the high energy emitting region along the kpc jet.
		 }  
 \end{figure}

\section{M87: A Toy Model}
\label{sec:M87}

Unified structure for the inner regions of quasars  \citep{1995PASP..107..803U,2000ApJ...545...63E}
demonstrate that the Active Galactic Nuclei (AGNs) all have similar internal structure.
The fraction of quasars with strong radio emission constitute 
about $\sim$10\% of the sample \citep{2002AJ....124.2364I,2014arXiv1401.1535K}.
The differences in their appearance largely result from the orientation of the jet.
When the jet is pointed close to the line-of-sight,  the emission is relativistically boosted.
The resulting extremely luminous object is called a blazar. 

At the center of the blazar,  the super massive black hole  has a typical Schwarzschild radius 
of $\sim 10^{-4}\,$ pc. 
So far, primary constraints on the physical scale of the central engines of distant 
AGN come from the intrinsic variability of the emission \citep{2001sac..conf....3P} and from  microlensing \citep{2010ApJ...712.1129M}. 
The jets transport energy and momentum over large distances \citep{1974MNRAS.169..395B}.
The jet may extend for scales up to megaparsecs.  Their structure is complex and includes
hotspots and blobs \citep{2011ApJS..197...24M,2012ApJ...755..174G,2012ApJS..203...31M}.

Most of the time, blazars are in the so-called low state; occasionally their flux increases by more than an order of magnitude. 
The rapid variability in these GeV-TeV flares implies 
that emission during a flare must originate from compact structures within the relativistic jet   \citep{2013ApJ...768...54B,2008MNRAS.384L..19B}.
It is increasingly clear that there may be a multiplicity of active zones during both the low and high 
states of blazars. 

3C273 provides an interesting local case in  point. 
The source is a typical blazar with the jet oriented nearly along the line-of-sight. 
Radio, optical and X-ray observations show that there is rich structure within the jet; 
the compact structures are different at different wavelengths \citep{2001ApJ...549L.167M}. 
VHE\footnote{very high energy, VHE: $>$ 100 GeV}  emission has not yet been detected from 3C273.

To understand the role of strong lensing in placing constraints on blazar structure, 
we use the detailed observations of M87 to construct a toy model.
M87 is a giant radio galaxy located in the core of the Virgo cluster at distance of 16.7 Mpc \citep{2005ApJ...634.1002J}. 
M87 has a blazar-like core and a jet, but it has lower luminosity than the typical high redshift blazar. 
M87 is  a VHE $\gamma$-ray emitter  \citep{2012ApJ...746..151A}. 
The exact location of the  source of VHE emission remains unclear
but the {\it Chandra X-Ray Observatory} monitoring program revealed increased intensity by more than a factor of 50 from the knot HST-1 \citep{2006ApJ...640..211H,2009ApJ...699..305H}.
In spite of the lower luminosity, we can use this well-observed nearby object as a platform for examining the power of strong lensing observations.

The M87 jet consists of bright knots of radio, optical, and X-ray emission spread throughout a projected distance of 1.6 kpc  (81 pc/").
The optical and X-ray emission correspond to radio features. 

Observations with the {\it Chandra X-Ray
Observatory} \citep{2007Ap&SS.311..329M} show that substantially  increased X-ray emission originates from at least two regions:
the core of M87 and a bright blob of material, HST-1, embedded in the jet \citep{1999ApJ...520..621B}. 
VERITAS, H.E.S.S. and MAGIC all detected VHE emission simultaneous with the increased X-ray luminosity from the HST-1.

The knot, HST-1, is located at a projected distance of $\sim$60~pc  from the core.
It consists of bright knots observed with apparent velocities, $\beta_{app}$, with a range of 4c-6c \citep{1999ApJ...520..621B}.
The observed apparent velocity requires that the jet must be no more than 19$^\circ$ from our line-of-sight~\citep{1999ApJ...520..621B} based on the standard picture of relativistic boosting \citep{1966Natur.211..468R}.

At the distance of M87,  the angular resolution of ground-based VHE instruments corresponds to  
a projected size of 30~kpc, much larger than the projected distance between the core and HST-1,  
thus explaining the ambiguity in the location of the VHE emission. 
Obviously, in the case of distant blazars, 
direct determination of the location of the source of the VHE emission is hopeless with current technologies.

In the discussion below we place M87 at redshift $z \sim 1$ with a lens at $z \sim 0.6$. 
We assume that the orientation of the M87 jet relative to the line-of-sight is the same as observed at its true redshift. 
We use this model to illustrate the way strong lensing can constrain the location of the VHE emission.
For the purposes of demonstrating the technique, we orient the jet along a radial direction in the lens plane.
%

\section{Physical Scales}
\label{sec:Scales}

As a first step in investigating strong lensing of blazars, we examine the range of projected physical scales relative to the Einstein radius in the lens plane.

We model the lens as a singular isothermal sphere (SIS) with Einstein radius:
\begin{equation}
\theta_E = \frac{r_E}{D_{OL}} = 4\pi \frac{\sigma^2}{c^2} \frac{D_{LS}}{D_{OS}} \,,
\label{eq:thetaE}
\end{equation} 
where $\sigma$ is the velocity dispersion of the stars, $D_{OL}$, $D_{LS}$, and $D_{OS}$ (see Figure~\ref{fig:sketch})
are the angular diameter distances from the observer to the lens, from the lens to the 
source and from the observer to the source, respectively \citep{1996astro.ph..6001N}.  

For strong lensing, the typical mass of the lensing galaxy is  $\sim 10^{11}\,\mbox{M}_{\odot}$. 
Using equation~(\ref{eq:thetaE}), the size of the source projected into the lens plane is:
\begin{eqnarray}
S &=& r_E\times \frac{D_{OS}}{D_{OL}} 
\nonumber \\ &\approx&
1\,\mbox{kpc} \, \frac{D_{OS}}{D_{OL}} 
\left(
\frac{D_{OL}D_{LS}}{D_{OS}\,1\,\mbox{Gpc}} \right)^{\frac{1}{2}} \left( \frac{M}{10^{11}
  \,\mbox{M}_{\odot}} \right)^{\frac{1}{2}}\,.
\end{eqnarray}

For our M87 toy model, the Einstein radius is 0.45$^{\prime\prime}$. 
The projected distance between the core and HST-1,
corresponds to 0.027$\,r_E$. There are many knots at greater distances 
that even extend beyond the Einstein radius. In principle, any of these knots may be sources of
$\gamma$-ray emission.

\section{The Application of Strong Lensing}
\label{sec:Application}

Here we outline the steps in interpreting strong lensing observations of blazars. 
The approach takes advantage of multi-wavelength observations. 
Resolved images in the radio, optical, and X-ray are a foundation for understanding the origin of $\gamma$-ray emission.

Figure~\ref{fig:sketch} shows a schematic for the proposed approach.
Radio and optical instruments already  provide resolved images of dozens of gravitationally lensed flat spectrum radio quasars (FSRQs)  
(JVAC/CLASS\footnote{http://www.jb.man.ac.uk/research/gravlens/lensarch/lens.html}). 
For the FSRQs, the redshift of the source and lens are known.
Cosmological parameters like Hubble constant  are very-well determined by other observations 
\citep{2013arXiv1303.5076P}; we thus assume these precise values to determine the distances in the lens
configuration.

The image morphology together with the distances constrain the central mass density of the lens.
For a  source at $z \sim 1$ the lens is at low enough redshift that direct optical imaging and spectroscopy can also provide important constraints on the central mass density and mass distribution of the lens \citep{2008ApJ...684..248B,1998ApJ...509..561K}. In the following discussion we assume that the lens properties are known.

Even at wavelengths where the lensed image is resolved, the magnification ratios and time delays may not be identical at different wavelengths. In general, 
these differences have been interpreted as a result solely of microlensing \citep{2001ApJ...551..929O}. 
However, the example of M87 suggests that differences in the location 
and size of the emitting regions at different wavelengths may contribute substantially to the variation in properties of the lensed images. 

The optical observations of M87 shows extended emission on a scale of $\sim$1~kpc dominated by the host galaxy. In contrast, 
VLA observations at wavelengths 2 cm and 90~cm  reveal very different emission region sizes ranging from sub parsec to hundred kiloparsec scales, respectively. The emission regions include
a number of knots, filaments, and more extended structures. 

Structures in the M87 jet are also resolved in the X-ray. 
These sources are spatially coincident with emission sources in the UV and the radio. 
If M87 were farther away or if it were observed with poorer angular resolution, 
the X-ray  variability that we observe arising from a knot in the jet 
might well be ascribed to an emission source near the core \citep{2003ApJ...586L..41H}.

Constraints on the distribution of the emitting regions can be also obtained in 
the $\gamma$-ray regime even though the images are unresolved. 
The duration of flares at high energies is short ($\sim$2~days) compared with the time delays (dozen to hundreds of days). 
Thus individual flares that correspond to multiple lensed images can be resolved. 
Identification of the components of the flare provide both amplitudes and time delays. 
Thus we have a measure of the magnification ratio even though the source is unresolved.

We have a set of constraints on the locations and sizes of emitting regions at resolved wavelengths. We can then use the $\gamma$-ray observations to ask whether the emission originates from the same regions as the lower energy radiation or not. 

Next, in Section~\ref{sec:VarR} we quantify the variation in the magnification ratio.
Then, in Section \ref{sec:VarDT} we discuss the variation in the time delay.

\section{Variation in the Magnification Ratio}
\label{sec:VarR} 

In the simple case where there are two lensed images, 
the magnification ratio depends on the displacement between the source and the lens in the lens plane. 
The magnification ratio also depends on the mass distribution of the lens which we assume is known from other observations. 
For the purpose of demonstration, we continue here with the assumption of 
a radially symmetric lens represented by the SIS model \citep{1996astro.ph..6001N}. 

For a source that lies within the Einstein ring $r_S < r_E$,
where $r_S$ is the distance between the source and the weighted center of the mass in the lens plane,
multiple images appear at distances $u_{\pm}$ from the lens:
\begin{equation}
u_{\pm}=u\pm 1 \,,
\end{equation}
where $u$ is in dimensionless units normalized to the Einstein radius ($u=r_S/r_E$).

The magnification for images of a compact source is
\begin{equation}
A_{\pm}=1\pm u_{\pm}^{-1}\,.
\label{eq:psmag}
\end{equation}
The magnification ratio between the two images of the compact source is
\begin{equation}
\frac{A_+}{A_-}=\frac{u+1}{u-1}\,.
\label{eq:ratio}
\end{equation}
Because gravitational lensing conserves surface brightness, 
the magnification for  an extended sources is simply  the ratio 
of the image area to the source area \citep{1993ApJ...413L...7S}. 
Correspondingly, the image magnification of an extended  source, $A'_{\pm}$, is the integral of the source intensity $I(x,y)$ 
weighted by the point source magnification: 
\begin{equation}
A'_{\pm}=\frac{\int I(x,y)A_{\pm}(r)dxdy}{\int I(x,y)dx dy}\,,
\end{equation}
where $x$ and $y$ are defined by  $r=\sqrt{x^2+y^2}$. 
We approximate the surface brightness of the source with a gaussian distribution.  

 \begin{figure}
 \includegraphics[width=8.5cm,angle=-90]{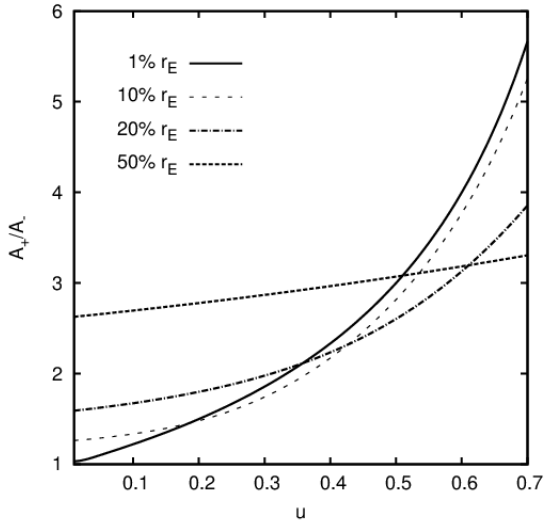}%
 \caption{\label{fig:ratioES} Magnification ratio between lensed images as 
               a function of the distance between the source and the lens in the lens plane 
               (in units of the Einstein radius).  
               The solid line corresponds to the magnification ratio of the images of the compact source.  
               Dotted, dashed-dotted and dashed lines show the magnification ratios 
               of images for an extended source of size 10\%, 20\% and 50\% $r_E$, respectively.
               We approximate the lens mass distribution with a SIS 
               and the surface brightness of the source with a gaussian distribution. }  
 \end{figure}

Figure~\ref{fig:ratioES} shows the magnification ratio between the images 
as a function of the distance between the lens and the source in the lens plane. 
The solid curve shows the magnification ratio for a compact source (size $\lesssim$1\%$r_E$; 
the dotted, dashed-dotted and dashed curves show the 
magnification ratio for an extended source with sizes 10\%, 20\%, 50\%  in units of $r_E$. 
Even a small change of the distance $u$ introduces  significant variation in the magnification ratio. 

Furthermore, concentric sources of emission at different wavelengths have different spatial extent and thus different magnification ratios. For example,
the majority of optical emission of M87, dominated by the host galaxy, would extends up to $\sim$50\% of the Einstein radius at z=1.
For our toy  model, the observed magnification ratio between the two optical images of the galaxy would be $\sim$3. Because the source is the extended host galaxy, the magnification ratio
would not change significantly with changes in $u$. 

For the M87 jet located at z=1, the distance between the core and the HST-1
corresponds to 0.027$\,r_E$. In the case of M87 the lensed image of these bright knots could be observed at radio and X-ray wavelengths. 
In the optical, their surface brightness is not high enough for them to be resolved against the entire galaxy. In the radio, the hot spots corresponding to the core and HST-1 are resolved. A change in the
distance between the lens and the source 
of the order of 0.027$\,r_E$ would correspond to a difference in the magnification ratio of  $\sim$0.2.
If the $\gamma$-ray source coincides with the radio source, the magnification should be identical.

For the M87 toy model, a 1\% $r_E$ change in $u$ corresponds to 22~pc.
During a flare, the large number of photons enables measurement of differences in the magnification ratio as small as  $\sim$0.2 and thus, differences in the location of the emission sites as small as 60 pc can be detected. 

Measurements of flux with even greater precision will allow  detection of smaller changes in the magnification ratio thus better discriminating among  sites of $\gamma$-ray emission.
Even with improved flux measurements, the magnification ratios alone are insufficient to restrict the location of $\gamma$-ray emission. As we discuss next, flaring of the sources and measurement of the time delays provides a route to a robust restriction of the source position.

\section{Variation in the Time Delay}
\label{sec:VarDT}

We can reduce the ambiguity in the origin of the $\gamma$-ray emission  by taking advantage of 
measurements of the time delay in flaring sources.
Once the time delay is measured, the  magnification ratio (or relative flux of the delayed components of the flare) between the images can also be determined. 
The difference in time delay is the essential measure of differences in  the location of emitting regions. 
 
The time delay function is \citep{1996astro.ph..6001N}:

\begin{equation}
\frac{c\, \Delta t(\vec{\theta})}{(z_L+1)} =
\frac{D_{OL}D_{OS}}{D_{LS}}\left [  \frac{1}{2}(\vec{\theta} -\vec{\theta_S})^2 -\psi(\vec{\theta})\right ]  \,,
 \label{dt3}
\end{equation}
where $\theta=r/D_{OL}$,  $\theta_S=r_S/D_{OL}$  and $\psi(\theta)$ is the effective 
gravitational potential of the lens. 
The time delay is proportional to the square of the angular offset between $\theta$ and $\theta_S$,
resulting  from the difference in light travel time for the two images. 

For the M87 toy model,
a 1\% $r_E$ change in $u$ (1\% $r_E$ = 22~pc) 
produces a change in the time delay of $\sim0.6$~days. 
The displacement of $\sim$0.027$r_E$ between the core  of M87 and HST-1
changes the time delay by almost 2 days
and  the magnification ratio by $\sim 0.2$. We measure the relative fluxes, or equivalently the magnification ratio, between the flare and its delayed counterpart.
We can then compare the magnification ratio with resolved sources at other wavelengths. If the magnification ratios are the same, it is consistent to identify the location of the $\gamma$-ray flare with the resolved source. 

\section{Limitations of the Model} 

Constraints on the size and location of the $\gamma$-ray emitting region site could be limited in some cases by 
microlensing, or by $\gamma$ - $\gamma$ absorption.

Microlensing  by a star within the lensing galaxy can impact the observed amplitude of the
$\gamma$-ray emission. The important issue in evaluating the impact of microlensing is
the Einstein radius of the microlensing star relative to the expected source size.
The Einstein radius of a star located at a cosmological distance is of the order of  $5\times 10^{16}$ ~(M/M$_{\odot}$)~cm. 
Limits on the size of the emitting region, obtained from the minimum variability time scale,  
are  also of the order of 10$^{16}$~cm \citep{2011AdSpR..48..998S}.  
\citet{HujCheung} provides an additional limit on the size of the $\gamma$-ray emitting region; 
they propose that the scale is $6\times 10^{14}$~cm times the Doppler factor. 
The average Doppler factor measured for $\gamma$-ray bright blazars is ~20  \citep{2010A&A...512A..24S} 
and thus the size limit is $1.2\times 10^{16}$~cm in rough accord with \citet{2011AdSpR..48..998S}.
The $\gamma$-ray emitting region is thus comparable in size to the Einstein radius of a solar-mass star. 

When the source size and Einstein radius of the lens are comparable, finite source effects
are important in microlensing \citep{2004ApJ...605...58K,2007MNRAS.376..263C,2003ApJ...584..657S}. 
In this case,  the magnification ratio is essentially unchanged. 

The upper limits on the sizes of emitting regions obtained for samples of 
blazars with variability time scales on the order of a few minutes
\citep{1996Natur.383..319G,2007ApJ...664L..71A,2011ApJ...730L...8A}
suggest that in some cases the size of the emitting region  can be smaller  compared to the Einstein radius. 
In these cases, the magnification ratio can differ from those predicted by the macro-lens modeling. 
Nevertheless, microlensing leaves the time delay unchanged. 
Thus, the location of the very compact $\gamma$-ray emitting regions can still be constrained  
using the proposed approach limited to time delay measurements.  

Another issue is that $\gamma$-ray photons emitted from sources at cosmological distances 
may be absorbed by the 
$\gamma$ - $\gamma$ interaction as they travel through various photon fields. 

\citet{2014arXiv1404.4422B} investigate whether  light emitted by the foreground lens affects 
the lensing signatures as a result of   $\gamma - \gamma$ absorption. 
They  find that the collective photon fields from lensing galaxies typically 
do not  produce any measurable excess $\gamma - \gamma$ opacity 
beyond that of the Extragalactic Background Light (EBL)  \citep{2010ApJ...723.1082A,2012Sci...338.1190A}.
The EBL can reduce the observed flux of the images of the lensed blazar, 
but the images of a given blazar will be changed by the same fractional amount. 
Thus, the magnification ratios and the time delays between the images remain unchanged by the EBL.

We conclude that neither microlensing nor absorption precludes application of 
the technique we propose for constraining the nature of the $\gamma$-ray source. 
In any case, once there is a large sample of lensed sources, 
it is unlikely that microlensing and/or absorption will have a significant effect on all sources.

\section{Discussion} 

Observation of gravitational lensing is challenging at high energies 
because only the most powerful sources, like blazars, can be detected. 
The nature of the jets in these sources remains puzzling.
Most of the power from  the majority of blazars is released in the high energy range, 
but investigating  high energy emissions from  these  sources 
is severely limited by the impossibility of resolving it. 
Despite abundant observations,
the origin of these $\gamma$-ray emissions is not fully understood. 
All models assume that the source of VHE emission is close to the base of the jet.  

The Fermi Gamma-ray Space Telescope has continuously monitored the whole sky since 2008, 
detecting photons from the most luminous objects in the universe (e.g. blazars and gamma-ray bursts).  
The number of expected gravitationally-lensed  systems among the 370 FSRQs  listed in the 2nd Fermi catalogue \citep{2012ApJS..199...31N}
is $\sim$10 \citep{2013arXiv1307.4050B},
but only about $\sim$5\% of these gravitationally-lensed systems will produce double images of the source \citep{1992ApJ...393....3F}.

The number of blazars detected is $\sim 1000$, but their redshifts are generally unknown. 
Thus, the number of lensed sources probably exceeds the estimate based
on the FSRQs. The lensing probability may be further  increased by  magnification  bias \citep{1984ApJ...284....1T,1993LIACo..31..217N}.
 
Among the FSRQs detected at high energy, PKS~1830-211  and B2~0218+35
are known as a strongly gravitationally lensed systems. 
The detection of a gravitational-lens induced echo of $27\pm 0.5$~days in the light curve of PKS~1830-211  during low state
by \citet{2011A&A...528L...3B} provided 
the first evidence for strong gravitational lensing of a
$\gamma$-ray source (see \citet{HujCheung} for an alternative view). 
Radio observations of the two lensed images of PKS~1830-211 
provide a  magnification ratio of $\sim$1.5 and time delay $26\pm4$~days \citep{1998ApJ...508L..51L}.
Remarkably, the delay for the $\gamma$-ray emission in the low state is consistent with the radio delay. 
Thus, the radiation sources may be coincident.
 
In contrast, the PKS~1830-211 flares detected in $\gamma$-rays \citep{2010ATel.2943....1C} 
do  not show delayed counterparts at the radio time delay. The magnification ratio has not yet been measured. A clean measurement of the time delay and the corresponding magnification ratio would
determine whether the flare and the emission in the low state originate from the  same region.
In other words, strong gravitational lensing can  be used to distinguish the sites of emission during low and flaring states of blazars
by comparing the time delay between delayed counterparts 
to the time delay obtained from the low state 
\citep{2011A&A...528L...3B}. This approach is   
similar to reverberation mapping \citep{1993PASP..105..247P}.

For the second lensed source, B2~0218+35, detected in high energies, the components of delayed $\gamma$-ray flare B2~0218+35 have
similar fluxes, implying a magnification ratio close to~1 \citep{HujCheung}.

\citet{HujCheung} demonstrates a method for analyzing complex light curves that 
consist of many superimposed flares. Their method allows extraction of both the time delay and 
the magnification ratios. 
Radio observations of this source give a time delay consistent with the time delay for the $\gamma$-ray flare. 
The time delay for the radio images at 15 and 8.4 GHz is $10.5\pm0.2$ days  \citep{1999MNRAS.304..349B} and $10.1\pm0.8$ days \citep{2000ApJ...545..578C}.
Radio observations  give much larger magnification ratios of 3.2$^{+0.3}_{-0.4}$ at 8~GHz  and 4.3$^{+0.5}_{-0.8}$ at 15~GHz \citep{2000ApJ...545..578C}. 

For B2~0218+35, one can construct a consistent model where the radiation sources coincide, but differ in size.
During the flare, the size of the $\gamma$ ray source is $<6\times10^{14}$~cm  \citep{HujCheung}, much smaller than the Einstein radius. 
In the radio, the jet extends well beyond the Einstein radius. 
Figure~\ref{fig:ratioES}  shows that the magnification ratio can easily differ by a factor of 3 for coincident sources with sizes $<1\%r_E$ and $\sim r_E$.  

For B2~0218+35, the large error in the radio time delay
precludes discrimination between the simple model of coincident radiation sources 
with different spatial extent and a model  similar to M87, 
where the radiation sources are not coincident.

If M87 were gravitationally lensed, and the flare originated from HST-1 rather than the core, 
we would expect a difference in the time delay of 
 $\sim$2~days. Robust discrimination between the core and HST-1 
 models would require an accuracy in time delay measurement of $\sim 0.4$ days. 
 This accuracy can probably be obtained in both the optical and $\gamma$-ray regimes.

The difference in magnification ratio for the core and HST-1 models is small, $\sim 0.2$. 
For a $\gamma$-ray flare, the error in the relative flux measurement can be $\lesssim 0.1$.  
These estimates assume a smooth, symmetric SIS mass distribution.
For more complex lenses with asymmetric mass distributions and possible substructure, 
variations of the magnification ratio and the time delay are larger. 
In principle, a more complex lens can enable the resolution of more detailed structure in the blazar.
Thus, our estimates for M87 are  conservative.

\citet{2010ApJ...722L...7P} suggests that the distances between the radio and $\gamma$-ray emissions are small.
Resolution of  small scales depends on the detailed properties of the lens. 
The inhomogeneous mass distribution of a typical galaxy allows  constraints on the projected distance 
to the $\gamma$-ray emitting region site that can be much tighter than 60 pc 
\citep{2002ApJ...567L...5M,2002ApJ...572...25D,2007MNRAS.378..109M,2009ApJ...699.1720K,2012MNRAS.419..936F}.


A model  where the VHE emission occurs close to the base of the jet and one where it originates from spatially distinct knots, as in M87, 
differ in the underlying fundamental physics~\citep{2003APh....18..593M,2004A&A...419...89R,2005ApJ...626..120S,2006MNRAS.370..981S}.
To explain the observed rapid variability and to avoid catastrophic pair production
in blazars, 
models assume that the $\gamma$-rays are produced in compact emission regions moving 
with relativistic bulk velocities in or near the parsec scale core \citep{1995MNRAS.273..583D}. 
However, recent detection of sub-TeV emission from FSRQs suggests 
that  the blazar zone can be located several parsecs away 
from a massive black hole \citep{2013arXiv1307.1779B,2012MNRAS.425.2519N}.
Refined analysis of existing data, along with additional identifications of gravitationally-lensed blazars,
have the potential to elucidate blazar emission models. 
 
\section{Conclusions}

Strong gravitational lensing is a potentially powerful tool for investigating the structures of jets at $\gamma$-ray energies
where the angular resolution of instruments is insufficient to  resolve the source. 
Observations of strongly lensed sources can discriminate among models where  the $\gamma$-ray emission comes
from the region close to the core and along those where it originates from knots along the jet.

Two gravitationally-lensed blazars have been detected at high energies
\citep{2011A&A...528L...3B,HujCheung}; both have flared \citep{2010ATel.2943....1C,2012ATel.4158....1C,2012ATel.4411....1C}.
Measurements of time delays and magnification ratios for the lensed counterparts 
provide a  unique opportunity to investigate the origin of the $\gamma$-ray emission.

The M87 toy model that we have constructed
shows that 
constraining the emission site  to a 60~pc projected distance requires 
 an accuracy in the time delay measurement of 0.4~days,
and  an accuracy in the  magnification ratio  of a few percent. 
Currently, these quantities can be  measured to this accuracy at $\gamma$-rays,
and in the near future  measurements with this precision will be possible  at shorter wavelengths. 

Shortly, dozens of gravitationally-lensed systems will be detected at energies $\lesssim 100$ GeV. 
In the coming years, these sources will have dozens of sets of flares. 
This statistical ensemble of flaring sources will allow a systematic study constraining the sites of $\gamma$-ray emission.

At energies $\gtrsim$ 100~GeV (VHE), absorption by the Extragalactic Background  Light 
precludes detection of sources at $z \gtrsim$ 1. 
Even at these energies, 
the number of known gravitationally-lensed systems  should increase very rapidly in the future. 
Euclid and SKA will reveal  blazars at redshift low enough to be detected at VHE.
Observations of flaring sources with ground-based Cherenkov telescopes like VERITAS \citep{2002APh....17..221W}, 
or, eventually, the Cherenkov Telescope Array, CTA \citep{2011ExA....32..193A}, 
will provide the time delays and magnification ratios necessary to limit the emission region. 

\acknowledgments
 
A.B. would like to thank Aneta Siemiginowska, Martin Elvis, Abraham Loeb, Markus B\"ottcher and Charles King
for comments and useful discussions.
We thank the referee for valuable comments on the manuscript.

\bibliography{resubmition}
\end{document}